\documentclass[11pt]{article}
\usepackage{epstopdf}
\usepackage{color}
\usepackage{subfigure}
\usepackage{amsmath}
\usepackage{amssymb}
\usepackage{graphicx,color}
\usepackage{cite}
\usepackage{enumerate}
\usepackage{amsthm}% theorems and proofs
\usepackage{amsfonts,mathrsfs}
\usepackage{geometry}
\usepackage{ifthen}
\usepackage{graphicx}
\usepackage{float}
\usepackage{bm}
\usepackage{booktabs}

\begin{document}
\begin{sloppypar}
\title{\bf Coherence monotones of quantum channels based on two generalized quantum relative entropies}
\vskip0.1in
\author{\small Jiaorui Fan$^1$, Zhaoqi Wu$^1$\thanks{Corresponding author. E-mail:
wuzhaoqi\_conquer@163.com}, Shao-Ming
Fei$^{2,3}$\\
{\small\it  1. Department of Mathematics, Nanchang University,
Nanchang 330031, China}\\
{\small\it  2. School of Mathematical Sciences, Capital Normal University, Beijing 100048,
China}\\
{\small\it  3. Max-Planck-Institute for Mathematics in the Sciences,
04103 Leipzig, Germany}}
\date{}
\maketitle

\noindent {\bf Abstract} {\small }\\
By using the Choi-Jamio{\l}kowski isomorphism, we propose two
classes of coherence monotones of quantum channels based on the
unified $(r,s)$-relative entropy and the sandwiched R\'{e}nyi relative
entropy. Elegant properties of the coherence monotones for quantum channels
are explored. Moreover, we present the upper bounds of the coherence monotones
and derive the explicit formulas of the coherence monotones for qubit unitary channels.\\


\vskip0.2in

\noindent {\bf 1. Introduction}\\\hspace*{\fill}\\
As a fundamental feature of quantum physics, coherence plays an
essential role in quantum information processing. Based on the
framework of quantifying the coherence of quantum states\cite{BC2014},
quantifications of quantum coherence have been extensively studied
in terms of the $l_{1}$-norm\cite{BC2014}, relative entropy\cite{BC2014},
skew information\cite{BA2018,Y2017,WZ2021}, fidelity\cite{U1976,LZ2017} and generalized
$\alpha$-$z$-relative R\'{e}nyi entropy\cite{ZJ2019}, with various
applications in quantum entanglement, quantum algorithm, quantum
meteorology and quantum
biology\cite{WZ2020,SS2015,DN20091
,DN20092,NB2016,JC2014,PH2008,L2011,LM2014,
MB2012,BD2015,L2005,MC2014}.

Quantum channels characterize the general evolutions of quantum
systems\cite{NC2010}. In recent years, fruitful results have been
obtained on studies of quantum channels
\cite{BK2017,ZS2017,WZ2022,XWF1,XWF2,
XWF3,H2016,KC2018,LY2020,VR2002}. Datta, Sazim, Pati and
Agrawal \cite{DS2018} investigated the coherence of quantum channels
by using the Choi-Jamio{\l}kowski isomorphism. Xu\cite{X2019}
proposed a framework to quantify the coherence of quantum channels
by using the Choi-Jamio{\l}kowski isomorphism, and defined the
$l_{1}$-norm coherence measure of quantum channels. Many quantifiers
of coherence for quantum channels have been put forward by employing
this framework, such as maximum relative entropy\cite{JY2021},
robustness\cite{JY2021}, fidelity\cite{WG2022}, skew information and
Hellinger distance\cite{XH2023}. Luo, Ye and Li\cite{LY2022}
introduced the coherence weight of quantum channels to investigate
the quantum resource theory of dynamical coherence. Kong, Wu, Lv,
Wang and Fei\cite{KW2022} presented an alternative framework to
quantify the coherence of quantum channels.

Entropy is a key concept in quantum information theory which
has many elegant properties. Hu and Ye\cite{HY2006} introduced the
quantum version of the unified $(r,s)$-entropy.
Wang, Wu and Minhyung\cite{WW2011} generalized
the unified $(r,s)$-entropy to the unified
$(r,s)$-relative entropy. Wilde, Winter and Yang\cite{WWY2014}
proposed the definition of the sandwiched R\'{e}nyi relative entropy.
Chitambar and Gour\cite{CG2016} defined the coherence monotone via the sandwiched
R\'{e}nyi relative entropy, and gave an elegant analytical expression
for pure states. Mu and Li\cite{ML2020} studied the quantum
uncertainty relations of the sandwiched R\'{e}nyi relative entropy of coherence
and the unified $(r,s)$-relative entropy of coherence.

The paper is organized as follows. In Section $2$, we give the
definition of coherence monotone of quantum channels based on the
unified $(r,s)$-relative entropy via Choi-Jamio{\l}kowski
isomorphism, and discuss several elegant properties. In
Section $3$, we study the coherence monotone of quantum channels
based on the sandwiched R\'{e}nyi relative entropy. In Section $4$,
we present explicit formulas of coherence monotones with respect to
qubit unitary channels for the two generalized quantum relative
entropies. Finally, we conclude in Section $5$.

\vskip0.1in

\noindent {\bf 2. Coherence monotone of quantum channels
based on the unified $\bm{(r,s)}$-relative entropy}\\\hspace*{\fill}\\
For two arbitrary quantum states $\rho$, $\sigma$ and $\alpha$,
$z$ $\in\mathbb{R}$, the unified $(r,s)$-relative entropy is defined by\cite{WW2011},
\begin{eqnarray}
\mathit{D}_{r}^{s}(\rho\parallel\sigma)=\left\{
\begin{array}{rcl}
H_{r}^{s}(\rho\parallel\sigma), & & \mathrm{if}~~0 \leq r < 1,~~s \neq 0,\\
H_{r}(\rho\parallel\sigma), & & \mathrm{if}~~0 \leq r < 1,~~s = 0,\\
H^{r}(\rho\parallel\sigma), & & \mathrm{if}~~0 \leq r < 1,~~s = 1,\\
_{\frac{1}{r}}{H(\rho\parallel\sigma)}, & & \mathrm{if}~~0 < r < 1,~~s = \frac{1}{r},\\
H(\rho\parallel\sigma), & & \mathrm{if}~~r = 1,
\end{array}\right.
\end{eqnarray}
where
\begin{align*}
&H_{r}^{s}(\rho\parallel\sigma)=
-[(1-r)s]^{-1}[(\mathrm{Tr}(\rho^{r}\sigma^{1-r}))^{s}-1],\\
&H_{r}(\rho\parallel\sigma)=-(1-r)^{-1}\mathrm{ln}(\mathrm{Tr}(\rho^{r}\sigma^{1-r})),\\
&H^{r}(\rho\parallel\sigma)=-(1-r)^{-1}[\mathrm{Tr}(\rho^{r}\sigma^{1-r})-1],\\
&
_{r}{H(\rho\parallel\sigma)}=
-(r-1)^{-1}[(\mathrm{Tr}(\rho^{\frac{1}{r}}\sigma^{1-\frac{1}{r}}))^{r}-1],\\
&H(\rho\parallel\sigma)=
\mathrm{Tr}(\rho\mathrm{ln}\rho)-\mathrm{Tr}(\rho\mathrm{ln}\sigma)
\end{align*}
are the quantum $(r,s)$-relative entropy, the quantum R\'{e}nyi
relative entropy of order $r$, the quantum Tsallis relative entropy,
the quantum relative entropy of type $r$ and the quantum
relative entropy, respectively.

Let $\{|i\rangle\}_{i=1}^{d}$ be a set of orthonormal basis of a
$d$-dimensional Hilbert space $H$. The set $\mathcal{I}$ of quantum
states is said to be incoherent if all the density matrices are
diagonal in this basis. The functional $\mathit{C}_{(r,s)}(\cdot)$
of a quantum state $\rho$ induced by the $(r,s)$-relative
entropy\cite{ML2020},
\begin{equation}\label{eq2}
\mathit{C}_{(r,s)}(\rho)=\mathop{\mathrm{min}}\limits_{\sigma\in
\mathcal{I}}\mathit{D}_{r}^{s}(\rho\parallel\sigma),
\end{equation}
is a coherence monotone when $r \in {(0,1)}$ and $s\leq1$.

Let $H_{A}$ and $H_{B}$ be two Hilbert spaces with dimensions $\vert
A\vert$ and $\vert B\vert$, orthonormal bases $\{|i\rangle\}_{i}$
and $\{|\beta\rangle\}_{\beta}$, respectively. Let
$\mathcal{D}(H_{A})$ and $\mathcal{D}(H_{B})$ be the set of all
density operators on $H_{A}$ and $H_{B}$, respectively.
%Assume that the orthonormal bases are fixed, and adopt the tensor basis $\{|i\beta\rangle\}_{i\beta}$ as the fixed bases with the multipartite system $H_{AB}=H_{A}\otimes H_{B}$.
Denote by $\mathcal{C}_{AB}$ the set of all channels from $\mathcal{D}(H_{A})$ to $\mathcal{D}(H_{B})$, $\mathcal{SC}_{ABA^{'}B^{'}}$ the set of all superchannels from $\mathcal{C}_{AB}$ to $\mathcal{C}_{A^{'}B^{'}}$, $\mathcal{IC}_{AB}$ the set of incoherent channels in $\mathcal{C}_{AB}$, and $\mathcal{ISC}_{ABA^{'}B^{'}}$ the set of incoherent superchannels in $\mathcal{SC}_{ABA^{'}B^{'}}$. A quantum channel $\phi\in{\mathcal{C}_{AB}}$ is a completely positive trace-preserving (CPTP) map. A coherence measure $\mathit{C}(\cdot)$ of quantum channels $\phi$ should satisfy the following conditions\cite{X2019}:\\
(a) Faithfulness: $\mathit{C}(\phi)\geq0$
for any $\phi\in{\mathcal{C}_{AB}}$, and $\mathit{C}(\phi)=0$ if and only if $\phi\in{\mathcal{IC}_{AB}}$;\\
(b) Nonincreasing under $\mathcal{ISC}s$:
$\mathit{C}(\phi)\geq\mathit{C[\mathrm{\Theta}(\phi)]}$ for any $\Theta\in{\mathcal{ISC}_{ABA^{'}B^{'}}}$;\\
(c) Nonincreasing under $\mathcal{ISC}s$ on average: $\mathit{C}\left(\phi\right)\geq\sum\limits_{m}p_{m}\mathit{C}(\phi_{m})$
for any $\Theta\in{\mathcal{ISC}_{ABA^{'}B^{'}}}$,
with $\{K_{m}\}_{m}$ an incoherent expression of $\Theta$, $p_{m}=\frac{\mathrm{Tr}(K_{m}J_{\phi}K_{m}^{\dagger})}{\vert{A^{'}}\vert}$ and $J_{\phi_{m}}=\vert{A^{'}}
\vert\frac{K_{m}J_{\phi}K_{m}^{\dagger}}{\mathrm{Tr}(K_{m}J_{\phi}K_{m}^{\dagger})}$;\\
(d) Convexity: $\mathit{C}\left(\sum\limits_{m}p_{m}\phi_{m}\right)
\leq\sum\limits_{m}p_{m}\mathit{C}(\phi_{m})$ for any
$\{\phi_{m}\}_{m}\subset\mathcal{C}_{AB}$ and probability
$\{p_{m}\}_{m}$.\\
Similar to the case of quantum coherence of quantum
states, we call the functional $C(\cdot)$ a coherence monotone of
quantum channels, if it satisfies conditions (a), (b)
and (d).\\\hspace*{\fill}\\
\noindent {\bf Definition 1} The unified $(r,s)$-relative entropy of
two arbitrary quantum channels $\phi$, $\widetilde{\phi}$
$\in\mathcal{C}_{AB}$ is defined by
\begin{eqnarray}
\mathit{D}_{r}^{s}(\mathit{M}_{\phi}\parallel
\mathit{M}_{\widetilde{\phi}})=\left\{
\begin{array}{rcl}
H_{r}^{s}(\mathit{M}_{\phi}\parallel
\mathit{M}_{\widetilde{\phi}}), & & \mathrm{if}~~0 \leq r < 1,~~s \neq 0,\\
H_{r}(\mathit{M}_{\phi}\parallel\mathit{M}_{\widetilde{\phi}}), & & \mathrm{if}~~0 \leq r < 1,~~s = 0,\\
H^{r}(\mathit{M}_{\phi}\parallel\mathit{M}_{\widetilde{\phi}}), & & \mathrm{if}~~0 \leq r < 1,~~s = 1,\\
_{\frac{1}{r}}{H(\mathit{M}_{\phi}\parallel\mathit{M}_{\widetilde{\phi}})}, & & \mathrm{if}~~0 < r < 1,~~s = \frac{1}{r},\\
H(\mathit{M}_{\phi}\parallel\mathit{M}_{\widetilde{\phi}}), & & \mathrm{if}~~r = 1,
\end{array}\right.
\end{eqnarray}
where $\mathit{M}_{\phi}=\frac{J_{\phi}}{\vert A\vert}$, and
$J_{\phi}$ is the Choi matrix corresponding to $\phi$.
\vskip0.1in

\noindent {\bf Definition 2} The coherence monotone of a channel
$\phi$ induced by the unified $(r,s)$-relative entropy is defined by
\begin{equation}\label{eq4}
\mathit{C}_{(r,s)}(\phi)=\mathop{\mathrm{min}}_{{\widetilde{\phi}}\in \mathcal{IC}_{AB}}D_{r}^{s}(\phi\parallel\widetilde{\phi})
=\mathop{\mathrm{min}}_
{\mathit{M}_{\widetilde{\phi}}\in\mathcal{I}}
\mathit{D}_{r}^{s}(\mathit{M}_{\phi}\parallel\mathit{M}_{\widetilde{\phi}}).
\end{equation}
\vskip0.1in

\noindent {\bf Theorem 1} $\mathit{C}_{(r,s)}(\phi)$ defined in Eq. (\ref{eq4})
is a coherence monotone of quantum channels
when $r \in (0,1)$ and $s \leq 1$.\\\hspace*{\fill}\\
\textbf{Proof}~ By Theorem $3.1$ in \cite{WW2011}, it is easy to see
that $\mathit{C}_{(r,s)}(\phi)\geq0$, and
$\mathit{C}_{(r,s)}(\phi)=0$ if and only if $\phi=\widetilde{\phi}$.
Thus, $\mathit{C}_{(r,s)}(\phi)$ satisfies (a).

Suppose
$\Theta^{'}=\frac{|A|}{|A^{'}|}\Theta$
with $\Theta\in{\mathcal{ISC}_{ABA^{'}B^{'}}}$. Thus, $J_{\Theta^{'}}$
is a CPTP map. We have
\begin{align*}
&D_{r}^{s}(J_{\Theta^{'}(\phi)}\parallel J_{\Theta^{'}(\widetilde{\phi})})\leq D_{r}^{s}(J_{\phi}\parallel J_{\widetilde{\phi}}).
\end{align*}
When $s\leq 0$, it can be found that
\begin{align*}
\left(\mathrm{Tr}\left(\left(\frac{J_{\Theta(\phi)}}{|A^{'}|}\right)^{r}
\left(\frac{J_{\Theta(\widetilde{\phi})}}{|A^{'}|}\right)^{1-r}\right)\right)^{s}
\leq \left(\mathrm{Tr}\left(\left(\frac{J_{\phi}}{|A|}\right)^{r}
\left(\frac{J_{\widetilde{\phi}}}{|A|}\right)^{1-r}\right)\right)^{s},
\end{align*}
which implies that
\begin{align*}
\frac{1}{(r-1)s}&\left[\left(\mathrm{Tr}
\left(\left(\frac{J_{\Theta(\phi)}}{|A^{'}|}\right)^{r}
\left(\frac{J_{\Theta(\widetilde{\phi})}}{|A^{'}|}\right)
^{1-r}\right)\right)^{s}-1\right]\\
\leq
&\frac{1}{(r-1)s}
\left[\left(\mathrm{Tr}\left(\left(\frac{J_{\phi}}{|A|}\right)^{r}
\left(\frac{J_{\widetilde{\phi}}}{|A|}\right)^{1-r}\right)\right)^{s}-1\right].
\end{align*}
By Eq. (\ref{eq2}), it follows that
\begin{align*}
&\mathit{D}_{r}^{s}\left(\frac{J_{\Theta(\phi)}}{|A^{'}|}\Big\|
\frac{J_{\Theta(\widetilde{\phi})}}{|A^{'}|}\right) \leq
\mathit{D}_{r}^{s}\left(\frac{J_{\phi}}{|A|}\Big\|
\frac{J_{\widetilde{\phi}}}{|A|}\right).
\end{align*}

Similarly, it can be seen that when $s \in(0,1] $,
\begin{align*}
\left(\mathrm{Tr}\left(\left(\frac{J_{\Theta(\phi)}}{|A^{'}|}\right)^{r}
\left(\frac{J_{\Theta(\widetilde{\phi})}}{|A^{'}|}\right)^{1-r}\right)\right)^{s}
\geq \left(\mathrm{Tr}\left(\left(\frac{J_{\phi}}{|A|}\right)^{r}
\left(\frac{J_{\widetilde{\phi}}}{|A|}\right)^{1-r}\right)\right)^{s},
\end{align*}
which implies that
\begin{align*}
\frac{1}{(r-1)s}&\left[\left(\mathrm{Tr}
\left(\left(\frac{J_{\Theta(\phi)}}{|A^{'}|}\right)^{r}
\left(\frac{J_{\Theta(\widetilde{\phi})}}{|A^{'}|}\right)
^{1-r}\right)\right)^{s}-1\right]\\
\leq
&\frac{1}{(r-1)s}\left
[\left(\mathrm{Tr}\left(\left(\frac{J_{\phi}}{|A|}\right)^{r}
\left(\frac{J_{\widetilde{\phi}}}{|A|}\right)^{1-r}\right)\right)
^{s}-1\right].
\end{align*}
According to Eq. (\ref{eq2}), it is obvious that
\begin{align*}
&\mathit{D}_{r}^{s}\left(\frac{J_{\Theta(\phi)}}{|A^{'}|}\Big\|
\frac{J_{\Theta(\widetilde{\phi})}}{|A^{'}|}\right) \leq
\mathit{D}_{r}^{s}\left(\frac{J_{\phi}}{|A|}\Big\|
\frac{J_{\widetilde{\phi}}}{|A|}\right).
\end{align*}
Therefore,
\begin{align*}
\mathit{C}_{(r,s)}(\Theta(\phi))&=\mathop{\mathrm{min}}
\limits_{\widetilde{\phi}\in{\mathcal{IC}_{AB}}}\mathit{D}_{r}^{s}
(\Theta(\phi)\parallel\widetilde{\phi})\\
&\leq \mathop{\mathrm{min}}\limits_{\widetilde{\phi}
\in{\mathcal{IC}_{AB}}}\mathit{D}_{r}^{s}
(\Theta(\phi)\parallel\Theta(\widetilde{\phi}))\\
&\leq \mathop{\mathrm{min}}\limits_{\widetilde{\phi}
\in{\mathcal{IC}_{AB}}}
\mathit{D}_{r}^{s}(\phi\parallel\widetilde{\phi})\\
&=\mathit{C}_{(r,s)}(\phi).
\end{align*}
Hence, (b) follows immediately.

Suppose that
$\mathit{C}_{(r,s)}(M_{\phi_{n}})= D_{r}^{s}(M_{\phi_{n}}\parallel M_{\widetilde{\phi}_{n}}^{\ast})$. Then
\begin{align*}
\mathit{C}_{(r,s)}\left(\sum_{n}p_{n}M_{\phi_{n}}\right)
&=\mathop{\mathrm{min}}_
{\mathit{M}_{\widetilde{\phi}_{n}}\in\mathcal{I}}
\mathit{D}_{r}^{s}
\left(\sum_{n}p_{n}M_{\phi_{n}}\parallel\mathit{M}_{\widetilde{\phi}_{n}}\right)\\
&\leq \mathit{D}_{r}^{s}\left(\sum_{n}p_{n}M_{\phi_{n}}\parallel
\sum_{n}p_{n}M_{\widetilde{\phi}_{n}}^{\ast}\right)\\
&\leq\sum_{n}p_{n}\mathit{D}_{r}^{s}(M_{\phi_{n}}\parallel
M_{\widetilde{\phi}_{n}}^{\ast})\\
&=\sum_{n}p_{n}\mathit{C}_{(r,s)}(M_{\phi_{n}}),
\end{align*}
where the second inequality follows from the joint convexity of $\mathit{D}_{r}^{s}(\cdot\parallel\cdot)$.
It implies that $\mathit{C}_{(r,s)}(\phi)$ satisfies (d).
This completes the proof.$\hfill\qedsymbol$

We now derive the analytical expression of the coherence monotone
based on the unified $(r,s)$-relative entropy. Let $M_{\widetilde{\phi}}
=\sum\limits_{{i\beta}}p_{i\beta}|i\beta\rangle\langle i\beta|$.
Then we have
\begin{align}\label{eq5}
\mathit{C}_{(r,s)}(\phi)
=\mathop{\mathrm{min}}_{p_{i\beta}}\frac{1}{(r-1)s}
\left[\left(\sum\limits_{i\beta}p_{i\beta}^{1-r}
\langle i\beta|M_{\phi}^{r}|i\beta\rangle\right)
^{s}-1\right].
\end{align}
Let $t_{i\beta}=\frac{\langle i\beta|M_{\phi}^{r}|i\beta\rangle}{t}$,
where $t=\sum\limits_{i\beta}\langle
i\beta|M_{\phi}^{r}|i\beta\rangle^{\frac{1}{r}}$.
Direct calculation shows that
$\sum\limits_{i\beta}t_{i\beta}=1$ and $t_{i\beta}\geq 0$.
Therefore,
\begin{align}\label{eq6}
\mathit{C}_{(r,s)}(\phi)
&=\mathop{\mathrm{min}}_
{p_{i\beta}}\frac{1}{(r-1)s}
\left[\left(\sum\limits_{i\beta}p_{i\beta}
^{1-r}\langle i\beta|M_{\phi}^{r}|i\beta\rangle\right)
^{s}-1\right]\notag\\
&=\frac{1}{(r-1)s}(t^{rs}-1)+t^{rs}\mathop{\mathrm{min}}_{p_{i\beta}}
\mathit{D}_{r}^{s}\left({t_{i\beta}}\parallel{p_{i\beta}}\right)\notag\\
&=\frac{1}{(r-1)s}(t^{rs}-1).
\end{align}

\vskip0.1in

\noindent {\bf Theorem 2} $\mathit{C}_{(r,s)}(\phi)$
has the following properties:\\
(i) for each $r \in (0,1)$, $\mathit{C}_{(r,s)}(\phi)$
is monotone decreasing with respect to $s$.\\
(ii) for each $r \in (0,1)$, $\mathit{C}_{(r,s)}(\phi)$
is a concave function of $s$.\\
{\bf Proof}\\
(i) It is easy to see that
\begin{align*}
\frac{\mathrm{d \mathit{C_{(r,s)}}(\phi)}}{\mathrm{d} s}
&=\frac{1}{(1-r)s^{2}}(t^{rs}-1)+\frac{1}
{(r-1)s}(t^{rs}\mathrm{ln}t^{r})\\
&=\frac{t^{rs}-1-t^{rs}\mathrm{ln}t^{rs}}{(1-r)s^{2}}.
\end{align*}
Let $m=t^{rs}$ and consider $y(m)=m-1-m\mathrm{ln}m$. When $s \in
(0,1)$, we have $0 < m \leq  1$, which implies that $y^{'}>0$.
Therefore, $ y(m)\leq y(1)=0 $. When $s \leq 0$, we have $ m\geq1 $
and $y^{'}<0$. Therefore, $ y(m)\leq y(1)=0 $. We thus obtain that
$\frac{\mathrm{d \mathit{C_{(r,s)}}(\phi)}}{\mathrm{d} s} \leq 0$.\\
(ii) According to (i), we have
\begin{align*}
\frac{\mathrm{d^{2} \mathit{C_{(r,s)}}(\phi)}}{\mathrm{d} s^{2}}
=\frac{2(t^{rs}-1-t^{rs}\mathrm{ln}t^{rs})+t^{rs}\mathrm{ln}t^{rs}
\mathrm{ln}t^{rs}}{(1-r)s^{3}}.
\end{align*}
Let $m=t^{rs}$ and consider
$f(m)=2m-2-2m\mathrm{ln}m+m(\mathrm{ln}m)^{2}$. It's easy to find
that $f^{'}(m) \geq 0$. When $s\in(0,1]$, we have $m \in (0,1]$.
Therefore, $f(m)<f(1)=0$. When $s\leq0$, we have $m \geq 1$.
Therefore, $f(m)>f(1)=0$. We thus obtain that
\begin{align*}
\frac{\mathrm{d^{2} \mathit{C_{(r,s)}}(\phi)}}{\mathrm{d} s^{2}}<0.
\end{align*}
This completes the proof.$\hfill\qedsymbol$

\vskip0.1in

\noindent {\bf 3. Coherence of quantum channels based on the sandwiched R\'{e}nyi relative entropy}\\\hspace*{\fill}\\
For any $r \in {(\frac{1}{2},1)}\cup {(1,\infty]}$,
the sandwiched R\'{e}nyi relative entropy is given by\cite{WWY2014}
\begin{align}
\widetilde{\mathit{D}}_{r}(\rho\parallel\sigma)=
\frac{1}{r-1}\mathrm{log}_{2}
\mathrm{Tr}\left[\left(\sigma^{\frac{1-r}
{2r}}\rho\sigma^{\frac{1-r}{2r}}\right)
^{r}\right].
\end{align}
Utilizing this distance measure, the sandwiched R\'{e}nyi relative
entropy of coherence monotone is defined by\cite{CG2016}
\begin{align}
\widetilde{\mathit{C}}_{R}(\rho)=\mathop{\mathrm{min}}\limits_{\sigma\in \mathcal{I}}\widetilde{\mathit{D}}_{r}(\rho\parallel\sigma).
\end{align}
In particular, when $\rho=|\psi\rangle\langle\psi|$ is a pure state,
it follows that\cite{CG2016}
\begin{align}
\widetilde{\mathit{C}}_{R}(\rho)=
\frac{2r-1}{r-1}\mathrm{log}_{2}t,
\end{align}
where $t=\sum\limits_{i}
(\langle\psi|i\rangle\langle i|\psi\rangle)^{\frac{r}{2r-1}}$.

When $\rho$ is a mixed state and $r \in {(\frac{1}{2},1)}\cup {(1,\infty]}$,
the sandwiched R\'{e}nyi relative entropy of coherence monotone
is defined by the convex-roof extension\cite{ML2020}
\begin{align}
\widetilde{\mathit{C}}_{R}^{\amalg}(\rho)
=\mathrm{min}
\left\{\sum\limits_{n}\lambda_{n}\widetilde{\mathit{C}}_{R}(|\psi_{n}\rangle)
| \rho=\sum\limits_{n}\lambda_{n}|\psi_{n}\rangle\langle\psi_{n}|\right\},
\end{align}
where the minimum is taken over all possible pure-state decompositions of $\rho=\sum\limits_{n}\lambda_{n}|\psi_{n}\rangle\langle\psi_{n}|$.\\
\noindent {\bf Definition 3} The sandwiched R\'{e}nyi
relative entropy of two arbitrary quantum channels $\phi$,
$\widetilde{\phi}$ $\in\mathcal{C}_{AB}$ is defined as
\begin{align*}
\widetilde{\mathit{D}}_{r}(\phi\parallel\widetilde{\phi})=
\frac{1}{r-1}\mathrm{log}_{2}
\mathrm{Tr}\left[\left(M_{\widetilde{\phi}}^{\frac{1-r}{2r}}
M_{\phi} M_{\widetilde{\phi}}^{\frac{1-r}{2r}}\right)
^{r}\right].
\end{align*}
\vskip0.1in

\noindent {\bf Definition 4} For a channel $\phi$,
if the Choi-Jamio{\l}kowski state
$M_{\phi}=|\varphi\rangle\langle\varphi|$
is a pure state, then the coherence monotone of
$\phi$ induced by the sandwiched R\'{e}nyi relative entropy can be defined as
\begin{equation}\label{eq11}
\widetilde{\mathit{C}}_{(R)}(\phi)
=\frac{2r-1}{r-1}\mathrm{log}_{2}\widetilde{t},
\end{equation}
where $\widetilde{t}=\sum\limits_{i\beta}
(\langle\varphi|i\beta\rangle\langle i\beta|\varphi\rangle)^{\frac{r}{2r-1}}$.
For the general case, we define the sandwiched R\'{e}nyi relative entropy of coherence monotone
by the convex-roof extension
\begin{align}\label{eq12}
\widetilde{\mathit{C}}_{R}^{\amalg}(\phi)
=\mathrm{min}
\left\{\sum\limits_{n}\lambda_{n}\widetilde{\mathit{C}}_{R}(\phi_{n})
|\phi=\sum\limits_{n}\lambda_{n}\phi_{n}\right\},
\end{align}
where the minimum is taken over all possible pure-channel decompositions of $\phi=\sum\limits_{n}\lambda_{n}\phi_{n}$. For any $n$, $M_{\phi_{n}}$
is the pure Choi-Jamio{\l}kowski state corresponding to the $\phi_{n}$.
\vskip0.1in

\noindent {\bf Theorem 3} $\widetilde{\mathit{C}}_{R}^{\amalg}(\phi)$
defined in Eq. (\ref{eq12})
is a coherence monotone when $r \in {(\frac{1}{2},1)}\cup {(1,\infty]}$.\\\hspace*{\fill}\\
\textbf{Proof}~ Since $\widetilde{\mathit{C}}_{R}(\phi_{n})\geq 0$,
it follows that $\widetilde{\mathit{C}}_{R}^{\amalg}(\phi)\geq 0$.
When $\widetilde{\mathit{C}}_{R}^{\amalg}(\phi)= 0$, there is an
ensemble $\{\lambda_{n},\phi_{n}\}$ such that
$\sum\limits_{n}\lambda_{n}\widetilde{\mathit{C}}_{R}(\phi_{n})=0$.
Then $\widetilde{\mathit{C}}_{R}(\phi_{n})=0$ for any $n$. Hence,
$\phi_{n}\in{\mathcal{IC}_{AB}}$ and $\phi\in{\mathcal{IC}_{AB}}$.
Noting that $\phi_{n}\in{\mathcal{IC}_{AB}}$, one easily finds that
$\sum\limits_{n}\lambda_{n}\widetilde{\mathit{C}}_{R}(\phi_{n})=0$,
i.e., $\widetilde{\mathit{C}}_{R}^{\amalg}(\phi)=0$. Hence, (a) is
proved.

Suppose
$\Theta^{'}=\frac{|A|}{|A^{'}|}\Theta$ with $\Theta\in{\mathcal{ISC}_{ABA^{'}B^{'}}}$. Thus, $J_{\Theta^{'}}$ is a CPTP map. We have
\begin{align*}
\widetilde{\mathit{C}}_{R}
\left(\frac{|A|}{|A^{'}|}J_{\Theta(\phi)}\right)
\leq\widetilde{\mathit{C}}_{R}(J_{\phi}).
\end{align*}
It is not difficult to see that
\begin{align*}
\widetilde{\mathit{C}}_{R}
\left(\frac{J_{\Theta(\phi)}}{|A^{'}|}\right)
\leq\widetilde{\mathit{C}}_{R}\left(\frac{J_{\phi}}{|A|}\right).
\end{align*}
Let $\phi=\sum\limits_{n}\lambda_{n}\phi_{n}$ be the optimal decomposition for
$\widetilde{\mathit{C}}_{R}^{\amalg}
(\phi)$. Then we have
\begin{align*}
\widetilde{\mathit{C}}_{R}^{\amalg}
(\Theta(\phi))
&=\widetilde{\mathit{C}}_{R}^{\amalg}
\left(\frac{J_{\Theta(\phi)}}{|A^{'}|}\right)\\
&\leq
\sum\limits_{n}\lambda_{n}\widetilde{\mathit{C}}_{R}
\left(\frac{J_\Theta({\phi_{n}})}{|A|}\right)\\
&\leq\sum\limits_{n}\lambda_{n}\widetilde{\mathit{C}}_{R}\left(\frac{J_{\phi_{n}}}{|A|}\right)\\
&=
\sum\limits_{n}\lambda_{n}\widetilde{\mathit{C}}_{R}(\phi_{n})=
\widetilde{\mathit{C}}_{R}^{\amalg}
(\phi).
\end{align*}
Therefore, (b) follows immediately.

For any ensemble $\{p_{n},\phi_{n}\}$, let $\phi=\sum\limits_{n}p_{n}\phi_{n}$. Then
\begin{align*}
\widetilde{\mathit{C}}_{R}^{\amalg}(\phi)
&=\widetilde{\mathit{C}}_{R}^{\amalg}\left(\sum_{n}p_{n}\phi_{n}
\right)\\
&\leq \sum\limits_{n}p_{n}\widetilde{\mathit{C}}_{R}(\rho_{n})\\
&=\sum\limits_{n}p_{n}\widetilde{\mathit{C}}_{R}^{\amalg}(\rho_{n}).
\end{align*}
Therefore, $\widetilde{\mathit{C}}_{R}^{\amalg}(\phi)$ satisfies (d).
This completes the proof.$\hfill\qedsymbol$

\vskip0.1in

\noindent {\bf 4. Example}\\\hspace*{\fill}\\
In this section, we choose the unitary channel to calculate the
coherence monotones based on the two quantum relative
entropies.$\\\hspace*{\fill}\\$ \noindent {\bf  Example 1} Consider
the unitary channel $\phi_{U}$ induced by the general qubit unitary
operator,
\begin{align}
U=
\left(\begin{array}{cc}
a & -b\\
e^{2\mathrm{i}\alpha}\bar{b} & e^{2\mathrm{i}\alpha}\bar{a}\\
\end{array}\right),
\end{align}
where
$a=e^{\mathrm{i}\left(\alpha-\frac{\beta}{2}-\frac{\delta}{2}\cos\frac{\gamma}{2}\right)}$ and
$b=e^{\mathrm{i}\left(\alpha-\frac{\beta}{2}+\frac{\delta}{2}\sin\frac{\gamma}{2}\right)}$.
Here, $\alpha$, $\beta$, $\gamma$ and $\delta$ are any real numbers.

The Choi-Jamio{\l}kowski state corresponding to $\phi_{U}$ is
\begin{align*}
\mathit{M}_{\phi_{U}}
=&(\mathbb{I}\otimes U)\left(\frac{1}{2}\sum_{i,j=0}^{1}|ii\rangle\langle jj|\right)(\mathbb{I}\otimes U)^{\dagger}\\
=&\frac{1}{2}\left(\begin{array}{cccc}
a\bar{a} & e^{-2\mathrm{i}\alpha}ab & -a\bar{b} &e^{-2\mathrm{i}\alpha}aa\\
e^{2\mathrm{i}\alpha}\bar{a}\bar{b} & b\bar{b} & -e^{2\mathrm{i}\alpha}b\bar{b} & a\bar{b}\\
-\bar{a}b & -e^{-2\mathrm{i}\alpha}bb & b\bar{b} & -e^{-2\mathrm{i}\alpha}ab\\
e^{2\mathrm{i}\alpha}\bar{a}\bar{a} & \bar{a}b & -e^{2\mathrm{i}\alpha}\bar{a}\bar{b} & a\bar{a}\\\end{array}\right).
\end{align*}
By calculation, we obtain the coherence monotone of $\phi_{U}$ via
the unified $(r,s)$-relative entropy as
\begin{align}\label{eq14}
\mathit{C}_{(r,s)}({\phi_{U}})&=\frac{\left(\sum\limits_{i,\beta=0}^{1}\langle i\beta|\mathit{M}_{{\phi_{U}}}^{r}|i\beta\rangle^{\frac{1}{r}}\right)^{rs}-1}{(r-1)s}
=\frac{2^{rs-s}\left[\left(\cos^{2}\frac{\gamma}{2}\right)^{\frac{1}{r}}
+\left(\sin^{2}\frac{\gamma}{2}\right)^{\frac{1}{r}}\right]^{rs}-1}
{(r-1)s}.
\end{align}
Noting that $x^{\frac{1}{r}}$ is a convex function when $r \in
(0,1)$, we have
\begin{align*}
\left(\cos^{2}\frac{\gamma}{2}\right)^{\frac{1}{r}}+
\left(\sin^{2}\frac{\gamma}{2}\right)^{\frac{1}{r}}
\geq2^{\frac{r-1}{r}}>0,
\end{align*}
which implies that
\begin{align*}
\left[\left(\cos^{2}\frac{\gamma}{2}\right)^{\frac{1}{r}}
+\left(\sin^{2}\frac{\gamma}{2}\right)^{\frac{1}{r}}\right]^{rs}
\geq \left(2^{\frac{r-1}{r}}\right)^{rs},
\end{align*}
and thus
\begin{align*}
\frac{2^{rs-s}\left[
\left(\cos^{2}\frac{\gamma}{2}\right)^{\frac{1}{r}}
+\left(\sin^{2}\frac{\gamma}{2}\right)
^{\frac{1}{r}}\right]^{rs}-1}
{(r-1)s}
\leq\frac{4^{(r-1)s}-1}{(r-1)s}.
\end{align*}
Therefore,
$\mathit{C}_{(r,s)}({\phi_{U}}) \in
\left[0,\frac{4^{(r-1)s}-1}{(r-1)s}\right]$.

Since the Choi-Jamio{\l}kowski state $\mathit{M}_{\phi_{U}}$ is a
pure state, the coherence monotone of $\phi_{U}$ based on the
sandwiched R\'{e}nyi relative entropy is given by
\begin{align}\label{eq15}
\widetilde{\mathit{C}}_{R}(\phi_{U})=
\frac{2r-1}{r-1}\mathrm{log}_{2}\left[\left(\cos^{2}\frac{\gamma}{2}\right)
^{\frac{r}{2r-1}}+\left(\sin^{2}\frac{\gamma}{2}\right)^{\frac{r}{2r-1}}\right]
+1.
\end{align}
When $r \in [\frac{1}{2},1)$,
$x^{\frac{r}{2r-1}}$ is a convex function. So we have
\begin{align*}
\left(\cos^{2}\frac{\gamma}{2}\right)^{\frac{r}{2r-1}}+
\left(\sin^{2}\frac{\gamma}{2}\right)^{\frac{r}{2r-1}}\geq
\left(\frac{1}{2}\right)^{\frac{1-r}{2r-1}}
\left(\cos^{2}\frac{\gamma}{2}+\sin^{2}\frac{\gamma}{2}\right)
^{\frac{r}{2r-1}}
=2^{\frac{r-1}{2r-1}}.
\end{align*}
Noting that
\begin{align*}
\mathrm{log}_{2}\left[\left(\cos^{2}\frac{\gamma}{2}\right)
^{\frac{r}{2r-1}}+\left(\sin^{2}\frac{\gamma}{2}\right)^{\frac{r}{2r-1}}\right]
\geq \mathrm{log}_{2}2^{\frac{r-1}{2r-1}}=\frac{r-1}{2r-1},
\end{align*}
we have $\widetilde{\mathit{C}}_{R}(\phi_{U})\leq2$.\\
When $r \in (1,\infty)$, $x^{\frac{r}{2r-1}}$ is a concave function. So we have
\begin{align*}
\left(\cos^{2}\frac{\gamma}{2}\right)^{\frac{r}{2r-1}}+
\left(\sin^{2}\frac{\gamma}{2}\right)^{\frac{r}{2r-1}}\leq
\left(\frac{1}{2}\right)^{\frac{1-r}{2r-1}}
\left(\cos^{2}\frac{\gamma}{2}+\sin^{2}\frac{\gamma}{2}\right)
^{\frac{r}{2r-1}}
=2^{\frac{r-1}{2r-1}}.
\end{align*}
Therefore,
\begin{align*}
\mathrm{log}_{2}\left[\left(\cos^{2}\frac{\gamma}{2}\right)
^{\frac{r}{2r-1}}+\left(\sin^{2}\frac{\gamma}{2}\right)^{\frac{r}{2r-1}}\right]
\leq \mathrm{log}_{2}2^{\frac{r-1}{2r-1}}=\frac{r-1}{2r-1},
\end{align*}
which also yields that $\widetilde{\mathit{C}}_{R}(\phi_{U})\leq2$.
Hence, we obtain that $\widetilde{\mathit{C}}_{R}(\phi_{U})\in
[0,2]$ when $r \in {(\frac{1}{2},1)}\cup {(1,\infty]}$.

Now consider the special unitary channel $\phi_{H}$ induced by the
Hadamard gate $H$. The Choi-Jamio{\l}kowski state of $\phi_{H}$ is
\begin{align*}
\mathit{M}_{\phi_{H}}
=&(\mathbb{I}\otimes H)\left(\frac{1}{2}\sum_{i,j=0}^{1}|ii\rangle\langle jj|\right)(\mathbb{I}\otimes H)^{\dagger}\\
=&\frac{1}{4}\left(\begin{array}{cccc}
1 & 1 & 1 & -1\\ 1 & 1 & 1 & -1\\ 1 & 1 & 1 & -1\\-1 & -1 & -1 & 1\\\end{array}\right).
\end{align*}
By calculation, it can be easily seen that
\begin{align}\label{eq16}
\mathit{C}_{(r,s)}({\phi_{H}})&=\frac{\left(\sum\limits_{i,\beta=0}^{1}\langle
i\beta|\mathit{M}_{{\phi_{H}}}^{r}|i\beta\rangle^{\frac{1}{r}}\right)^{rs}-1}{(r-1)s}
=\frac{4^{(r-1)s}-1}{(r-1)s},
\end{align}
and
\begin{align}\label{eq17}
\widetilde{\mathit{C}}_{R}(\phi_{H})=
\frac{2r-1}{r-1}\mathrm{log}_{2}4^{\frac{r-1}{2r-1}}=2,
\end{align}
which means that the two quantifiers of coherence for unitary
channels induced by the Hadamard gate both reaches the maximum value
of the ones induced by the general qubit unitary operators.

\vskip0.1in

\noindent {\bf 5. Conclusions}\\\hspace*{\fill}\\
Following the idea in \cite{X2019}, we have introduced
the coherence monotones of quantum channels
by utilizing the Choi-Jamio{\l}kowski isomorphism
based on the unified $(r,s)$-relative entropy
and the sandwiched R\'{e}nyi relative entropy.
In addition, we have explored some elegant properties of
the coherence monotone via the unified $(r,s)$-relative entropy.
Furthermore, we have calculated the coherence monotones
of qubit unitary channels based on these two quantum relative
entropies. According to these results, we have also
studied the upper bounds of the coherence monotones for $\phi_{U}$.

\noindent

%=============================================================================%
\subsubsection*{Acknowledgements}
\small {This work was supported by National Natural Science
Foundation of China (Grant Nos. 12161056, 12075159, 12171044);
Beijing Natural Science Foundation (Grant No. Z190005); the Academician
Innovation Platform of Hainan Province.}
%===========================================================================%

%=============================================================================%
\subsubsection*{Conflict of interest}
\small {The authors declare that they have no conflict of interest.}

%===========================================================================%

%=============================================================================%
%===========================================================================%

\end{sloppypar}
\end{document}